\documentclass[preprintnumbers,
a4paper,
superscriptaddress,
prd,nofootinbib,floatfix,12pt,
]{revtex4}

\usepackage{graphicx,amssymb,amsmath,color}
\usepackage{hyperref}
\usepackage{color}
\usepackage{setspace}

\def\beq{\begin{equation}}
\def\eeq{\end{equation}}
\def\ceq{\end{equation} \begin{equation}}
\def\bea{\begin{eqnarray}}
\def\eea{\end{eqnarray}}
\def\bei{\begin{itemize}}
\def\eei{\end{itemize}}
\def\bmat{\begin{matrix}}
\def\emat{\end{matrix}}
\def\ble{\begin{flushleft}}
\def\ele{\end{flushleft}}
\def\={\,=\,}
\def\+{\,+\,}
\def\-{\,-\,}

\def\GeV{\,{\rm GeV}\,}

\newcommand{\Fig}[1]{Fig.~\ref{#1}}

\newcommand{\Sec}[1]{Sec.~\ref{#1}}

\begin{document}

\title{Implications of an axino LSP for naturalness}

\author{\vspace{2mm} Gabriela Barenboim}
\email{Gabriela.Barenboim@uv.es}
\affiliation{Departament de F\'{\i}sica Te\`orica and IFIC, Universitat de Val\`encia-CSIC, E-46100, Burjassot, Spain.}

\author{Eung Jin Chun}
\email{ejchun@kias.re.kr}
\affiliation{Korea Institute for Advanced Study, Seoul 130-722, Korea}
\affiliation{Kavli Institute for Theoretical Physics, University of California, Santa Barbara, CA 93106, USA}

\author{Sunghoon Jung}
\email{nejsh21@gmail.com}
\affiliation{Korea Institute for Advanced Study, Seoul 130-722, Korea}

\author{Wan Il Park}
\email{wipark@kias.re.kr}
\affiliation{Departament de F\'{\i}sica Te\`orica and IFIC, Universitat de Val\`encia-CSIC, E-46100, Burjassot, Spain.}
\affiliation{Korea Institute for Advanced Study, Seoul 130-722, Korea}

\begin{abstract} \vspace{3mm} \baselineskip=16pt
Both the naturalness of the electroweak symmetry breaking and the resolution of
the strong CP problem may require a small Higgsino mass $\mu$ generated by a realization of the DFSZ axion model. Assuming the axino is the lightest supersymmetric
particle, we study its implications on $\mu$ and the axion scale. 
Copiously produced
light Higgsinos at collider (effectively only neutral NLSP pairs) eventually decay to axinos leaving prompt multi-leptons or displaced vertices which are being looked for at the LHC. 
We use latest LHC7+8 results to derive current limits on $\mu$ and the axion scale. Various Higgsino-axino phenomenology
is illustrated by comparing with a standard case without  lightest axinos as well as
with a more general case with additional light gauginos in the spectrum.
\end{abstract}

\preprint{KIAS-P14037, NSF-KITP-14-086}

\maketitle

\newpage

\baselineskip=15pt
\tableofcontents

\baselineskip=18pt

\section{Introduction}

The strong CP problem is elegantly resolved by introducing a Peccei-Quinn (PQ) symmetry \cite{pq}
and its spontaneous breaking resulting in a dynamical field called the axion \cite{ww}. In this mechanism, the CP-violating
QCD $\theta$ term is determined by a vacuum expectation value of the axion which dynamically  {\color{black} cancels out the non-zero QCD $\theta$-term.}
The PQ symmetry can be realized  either by introducing heavy quarks (KSVZ) \cite{ksvz} or
by extending the Higgs sector (DFSZ) \cite{dfsz} and its  breaking breaking scale $v_{PQ}$ is related with the axion coupling constant as $f_a \equiv \sqrt{2} v_{\rm PQ}/N_{\rm DW}$ with $N_{\rm DW}$ being the domain wall number counting the QCD anomaly\footnote{The standard DFSZ model has $N_{DW}=6$, but a certain variations can allow $N_{DW}=1$ to avoid the domain wall problem}.
The conventionally allowed window of the axion coupling constant is $10^9 \lesssim f_a / \GeV \lesssim 10^{12}$
(For a review, see \cite{kim08}). The upper bound comes from the axion cold dark matter contribution which is cosmological model dependent. A recent simulation of axionic topological defect contributions provides a stringent upper bound $f_a / \GeV \lesssim \mbox{a few} \times 10^{10}$ if PQ-symmetry were broken after inflation \cite{axionwindow}.
The window can be widen if PQ-symmetry were broken before or during inflation in certain class of PQ symmetry breaking models avoiding too large axionic isocurvature perturbations \cite{highH}.
The existence of such a high scale causes quadratic divergences to the Higgs boson mass and thus requires a huge fine-tuning
to keep stable two scales, the electroweak scale and the PQ scale (or a generic UV scale).

Supersymmetry (SUSY) would be the best-known framework to avoid such a hierarchy problem.
However, the electroweak symmetry breaking in SUSY suffers from a certain degree of fine-tuning
to maintain a desirable potential minimization condition:
\begin{equation} \label{mZmin}
{m_Z^2 \over 2} = {m^2_{H_d} - m^2_{H_u} \tan^2\beta \over \tan^2\beta -1 } - \mu^2
\end{equation}
where $m_{H_{u,d}}$ are the soft masses of the two Higgs doublets, $\tan\beta \equiv v_u/v_d$ is the ratio of their vacuum
expectation values, and $\mu$ is the Higgs bilinear parameter in the superpotential.
As LHC finds no hint of SUSY, it pushes up
the soft mass scale above TeV range, the minimization condition (\ref{mZmin}) requires a fine cancellation among different terms.
Barring too huge cancellation, one may arrange $m_{H_{u,d}}$ and  $\mu$ not too larger than $m_Z$. This has been advocated as ``natural SUSY'' \cite{NS} implying stops/sbottoms at sub-TeV and light Higgsinos with
\begin{equation}
 \mu \lesssim 200 \,\mbox{GeV}.
\end{equation}
Such a spectrum can also be obtained radiatively with multi-TeV soft masses at a UV scale \cite{RNS}.

\medskip

An electroweak  $\mu$  may be related to the PQ symmetry in the manner of DFSZ \cite{kim84}, which introduces a non-renormalizable superpotential in the Higgs sector:
\begin{equation} \label{Wdfsz}
 W= \lambda_\mu {P^2 \over M_P} H_u H_d
\end{equation}
where $P$ and thus $H_u H_d$ carries a non-trivial PQ charge and $M_P$ is the reduced Planck mass.
Upon the PQ symmetry breaking $v_{PQ} \sim \langle P \rangle$,
a $\mu$ term is generated by $\mu = \lambda_\mu \langle P \rangle^2/M_P$.
{\color{black} Once PQ-symmetry is broken, there appear the axion $a$,  its scalar partner, the saxion $s$, and
the fermion super-partner, the axino $\tilde a$.}
Forming an axion superfield $A= (s+ ia, \tilde a)$, one can schematically write down
the effective $\mu$-term superpotential;
\begin{equation} \label{Weff}
 W = \mu H_u H_d + c_H {\mu \over v_{PQ} } A H_u H_d
\end{equation}
where $c_H$ is a parameter depending on the PQ symmetry breaking sector; we use $c_H=2$ in this paper.
In the context of the natural SUSY having a small $\mu$ parameter, a neutral Higgsino tends to be the lightest supersymmetric
particle (LSP) and thus is a dark matter candidate assuming R-parity. In this case, a heavy axino decay to the LSP can change
the standard thermal Higgsino dark matter density resulting in different mixtures of the axion and Higgsino dark matter components depending on the PQ scale \cite{bae13}.

\medskip

In this paper, we investigate implications of the axino LSP in the framework of ``the natural SUSY DFSZ model''.
Naively speaking, the axino mass is expected to be of order of the soft SUSY breaking scale, but it
is in general model dependent \cite{goto92,chun92}.
As a dark matter, the abundance of axino depends on the history of the universe involving either the condensation of saxion or the reheating temperature of the primordial inflation.
Axinos can be produced abundantly either by saxion decay \cite{chun00,Kim:2008yu}
or by interactions with thermal particles \cite{chun11,bae11,bae14,Park:2014qha}\footnote{For the axino dark matter property in the KSVZ model, see Ref.~\cite{choi13}.}. To avoid axino over-production, we assume the axino is very light or the reheat temperature is low enough to suppress the thermal production in this paper.

Since Higgsinos are predicted to be light in the natural SUSY scenario, they can be copiously produced at the LHC and decay to axino plus the Higgs boson $h$ or $Z$ boson through the coupling in Eq.~(\ref{Weff}).  This leads to interesting signatures of multi-leptons/jets and missing transverse energy(MET) which can be prompt or displaced depending on the PQ scale. Notice that the standard Higgsino LSP scenario is hard to probe as heavier Higgsino decays produce unobservably soft leptons or pions due to a small mass gap  between a heavier Higgsino and the Higgsino LSP.  Currently, the ATLAS and CMS collaborations look for prompt multi-lepton plus MET and displaced di-jet/lepton signatures. Applying the current search results to the Higgsino-axino system, we obtain various limits on the $\mu$ parameter as well as the PQ scale. We assume that sleptons, squarks and gluinos are heavy, but see Refs.~\cite{Brandenburg:2005he} for earlier collider studies in the presence of light sleptons.

In \Sec{sec:noaxino}, we first translate the current multi-lepton +MET search results to the Higgsino-bino system where the Higgsino and bino are
taken to be the next-to-LSP (NLSP) and the LSP, respectively, and thus the NLSP decay to the LSP plus $h$ or $Z$ can lead
to prompt multi-lepton signatures. In \Sec{sec:axinolsp}, we turn into a case of the Higgsino NLSP and the axino LSP which can lead to displaced vertices from the NLSP decay. Then, we extend our analysis to the case of the Higgsino NNLSP and the bino NLSP with the axino LSP in \Sec{sec:higgsinonnlsp}. 
LHC14 projections of displaced vertex searches are estimated in \Sec{sec:lhc14} to see how far the axion scale can be probed.
Finally, we conclude in \Sec{sec:conclusion}.

\section{Current limits on (N)LSP Higgsinos without axinos} \label{sec:noaxino}

\begin{figure}[t] \centering
\includegraphics[width=0.6\textwidth]{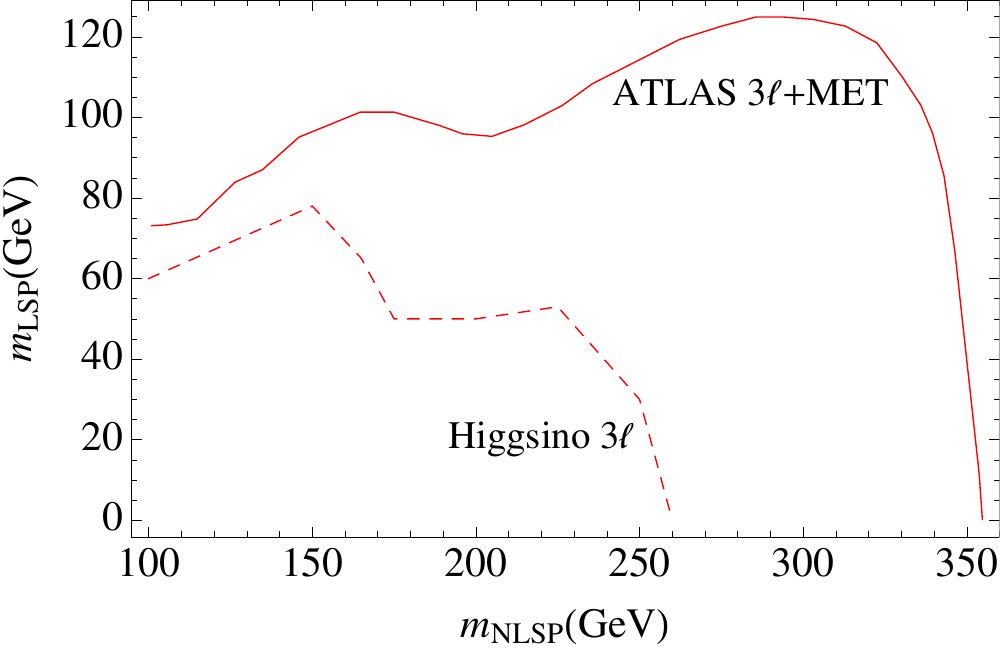}
\caption{\baselineskip=16pt
Current exclusion limits on the standard case with Higgino NLSP and Bino LSP. The official bound on the Wino NLSP (Bino LSP and 100\% branching ratios to $W$ and $Z$ bosons) from the $3\ell$+MET search (20.3/fb) is shown as the solid line for reference~\cite{Aad:2014nua}. Assuming the Higgsino NLSP with the Bino LSP, we re-interprete the $3\ell$+MET search(dashed).
The relevant BR is taken into account with $\mu>0$, and the $3\ell$ search is not sensitive to the sign of $\mu$ as depicted in \Fig{fig:BR-higgsinopair}. We assume $M_2=$2 TeV and $t_\beta=3$. More on \Sec{sec:noaxino}.}
\label{fig:higgsino3l}
\end{figure}

\begin{figure}[t] \centering
\includegraphics[width=0.49\textwidth]{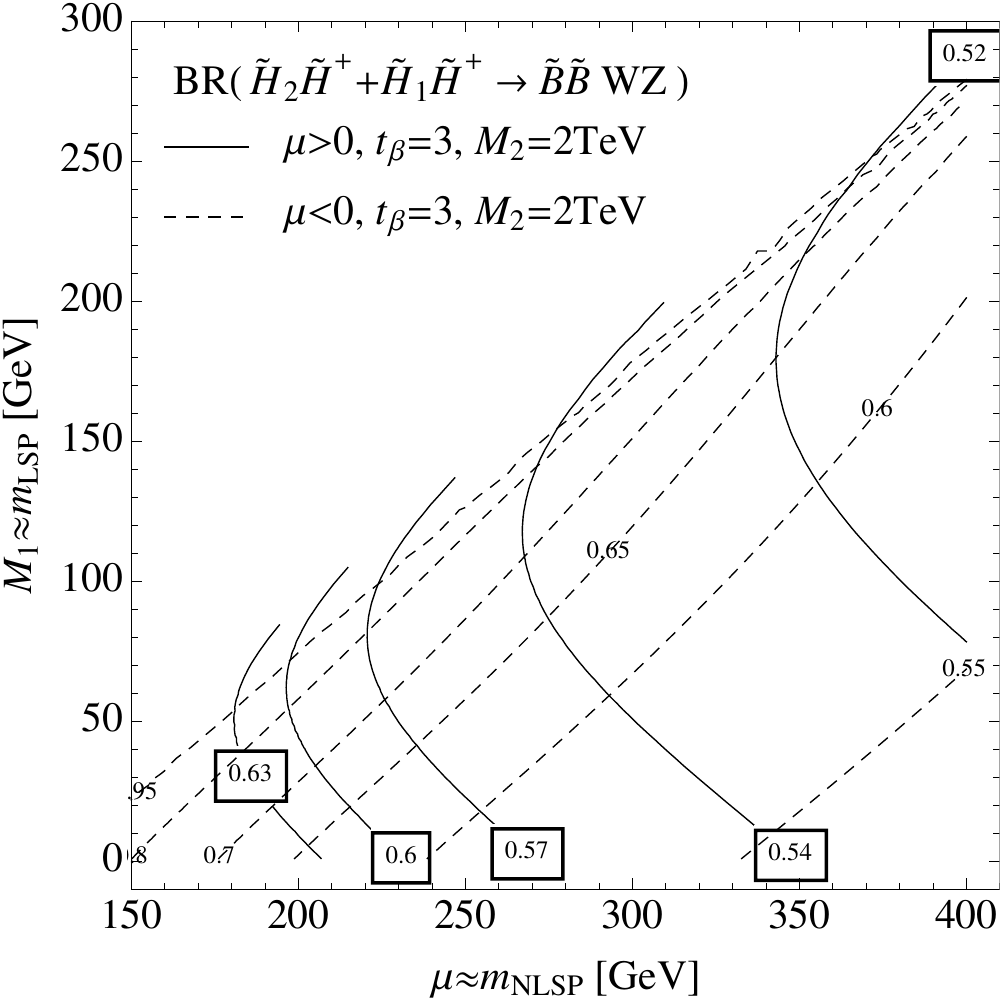}
\caption{\baselineskip=16pt
The branching ratios of NLSP Higgsino pairs to LSP Binos that are relevant to the $3\ell$+MET search.
Both $\mu>0$(solid) or $\mu<0$(dashed) are shown. The relevant BR shown does not vary much in most of the parameter space with both signs of $\mu$~\cite{Jung:2014bda}. This result is used in \Fig{fig:higgsino3l}.}
\label{fig:BR-higgsinopair}
\end{figure}

Before considering the axino LSP, let us first deduce and summarize the current exclusion bounds in the case of (1) the standard Higgsino-like NLSP and Bino-like LSP as well as in the case of (2) Higgsino-like LSP. The results will be later compared with those with axino LSPs. 

Consider first the case (1) with Higgsino NLSP and Bino LSP. Being relatively light, a sizable number of charged and neutral Higgisinos, $\chi_1^\pm$ and $\chi^0_{2,3}$, can be produced electroweakly and decay to the LSP $\chi^0_1$ through
 $\chi_1^\pm \to \chi_1^0 W^\pm$ and  $\chi_{2,3}^0 \to \chi_1^0 + h,Z$. The neutral Higgsino decays to the $Z$ boson are relevant to the multi-lepton searches\footnote{Contributions from intermediate Higgs bosons are generally small because of the small leptonic BR via $h \to WW^*, ZZ^*$ although Higgs decay products can certainly be useful when Higgsinos are heavy~\cite{Han:2013kza,Baer:2012ts,Jung:2014bda,Jung:future}; see also \cite{Ghosh:2012mc}. Considering light Higgsinos, we ignore Higgs contributions in this work.}, and its branching ratio (BR) is a function of $t_\beta$ and the sign of $\mu$. In \Fig{fig:higgsino3l}, we show current bounds on the Higgsino NLSP overlapping the officially reported bound on the Wino NLSPs from the $3\ell$+MET search~\cite{Aad:2014nua} for reference. The associate production of charged and neutral Higgsinos is the largest and is constrained from the $3\ell$+MET search: $\chi_1^\pm \chi_2^0 \to \chi_1^0 \chi_1^0 \, WZ \to \chi_1^0 \chi_1^0 + 3\ell \, \nu$. Here, the dependence on the underlying model parameters such as $t_\beta$ and the sign of $\mu$ is weak as demonstrated in the left panel of \Fig{fig:BR-higgsinopair} -- the relevant BR of Higgsino pairs is in general close to a half~\cite{Jung:2014bda}. So we use a positive $\mu$ to draw the bound in \Fig{fig:higgsino3l}. The bound on Higgsinos is weaker than the official bound on Winos due to two modifications:  i) The total production cross-section of Higgsino pair $\chi_2^0 \chi_1^\pm + \chi_3^0 \chi_1^\pm$ is smaller than that of Wino pairs $\chi_2^0 \chi_1^\pm$ (by about a factor 2 for ${\cal O}(100)$GeV NLSPs), and ii) the BR for $\chi^0_{2,3} \to \chi^0_1 + Z$ is smaller than 1. The actual bound on Winos will also be weaker than the officially reported one according to a smaller BR. On the other hand, other multi-lepton searches contributed mainly from other pair productions of Higgsinos currently lead to weaker or null bounds; for example, the associate production of two neutral Higgsinos is only weakly constrained from the $4\ell$+MET search~\cite{Khachatryan:2014qwa} via $\chi_2^0 \chi_3^0 \to \chi_1^0 \chi_1^0 \, ZZ \to \chi_1^0 \chi_1^0 \, 4\ell$. In all, the NLSP Higgsino mass exclusion currently reaches up to about 250GeV while  the LHC sensitivity drops quickly as the mass-gap between the NLSP and LSP becomes smaller.

 There exist other experimental results on $3\ell$+MET~\cite{Khachatryan:2014qwa} and $4\ell$ +MET~\cite{ATLAS:2013qla}. But they are not essentially different from the ones used above. The $2\ell + 2j$~\cite{Khachatryan:2014qwa,Aad:2014vma} and the same-sign dilepton~\cite{Khachatryan:2014qwa} searches do not give a much stronger bound for such a light $\mu$. The $2\ell + 0j + 0Z$ search for the $WW$ is also potentially useful~\cite{Aad:2014vma,Han:2013kza}. In any case, our interpretation of a few standard searches in \Fig{fig:higgsino3l} (and similar figures throughout in this paper) is a reasonable and useful estimation of current Higgsino exclusion limits. See Appendix~\ref{app:simulation} for more details on how we obtain the bounds.

The Higgsino can also be the LSP (the case (2) above). If other gauginos are far away in mass, all three Higgsino states -- one charged and two neutral -- are nearly degenerate. Even though light Higgsinos are abundantly produced, visible decay products of decays between them are generally too soft to be observable at collider and two LSPs are produced in back-to-back directions giving a small MET. It is why the search of nearly degenerate spectrum is difficult. The squeezed spectrum is typically searched by triggering hard initial state radiations(ISR) which subsequently boost the visible and invisible decay products. No dedicated LHC search is reported yet, but several theoretical studies of LHC prospects have been carried out in Refs.~\cite{Gori:2013ala}. It is expected that the monojet+MET alone at LHC14 would be sensitive to nearly degenerate $\sim 100$GeV Higgsinos only with ${\cal O}(1)$/ab of data, but somewhat more optimistic approach would be to utilize soft leptons from heavier Higgsino decays when the (model-dependent) mass-splitting is $\sim 10$GeV or larger. Decays between Higgsinos are rather prompt~\cite{Thomas:1998wy} (even when the splitting is dominated by small loop-induced contributions), so the disappearing track searches~\cite{TheATLAScollaboration:2013bia} that are sensitive to the degenerate Wino LPSs are not so useful for Higgsino LSPs; see Appendix~\ref{app:inodecays}.

\section{Higgsino NLSP and axino LSP} \label{sec:axinolsp}

\begin{figure} \centering
\includegraphics[width=0.49\textwidth]{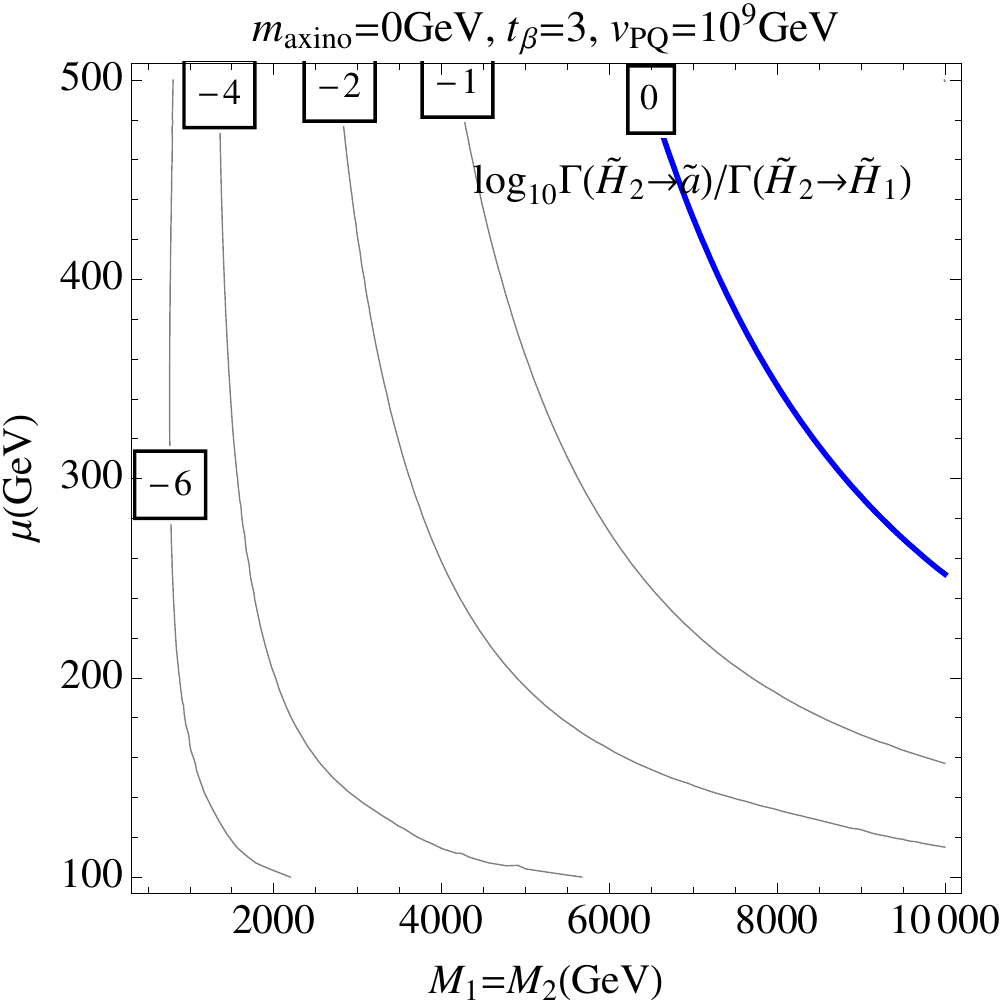}
\includegraphics[width=0.49\textwidth]{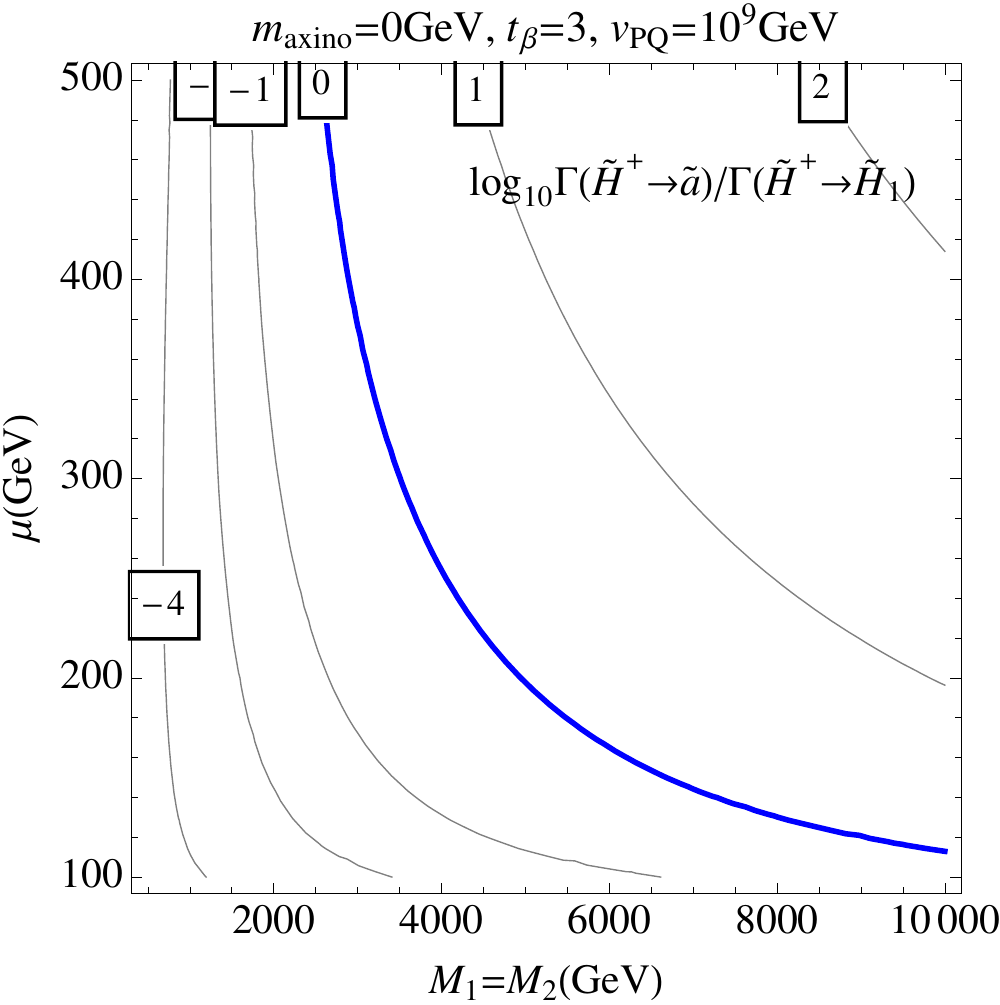}\\
\includegraphics[width=0.49\textwidth]{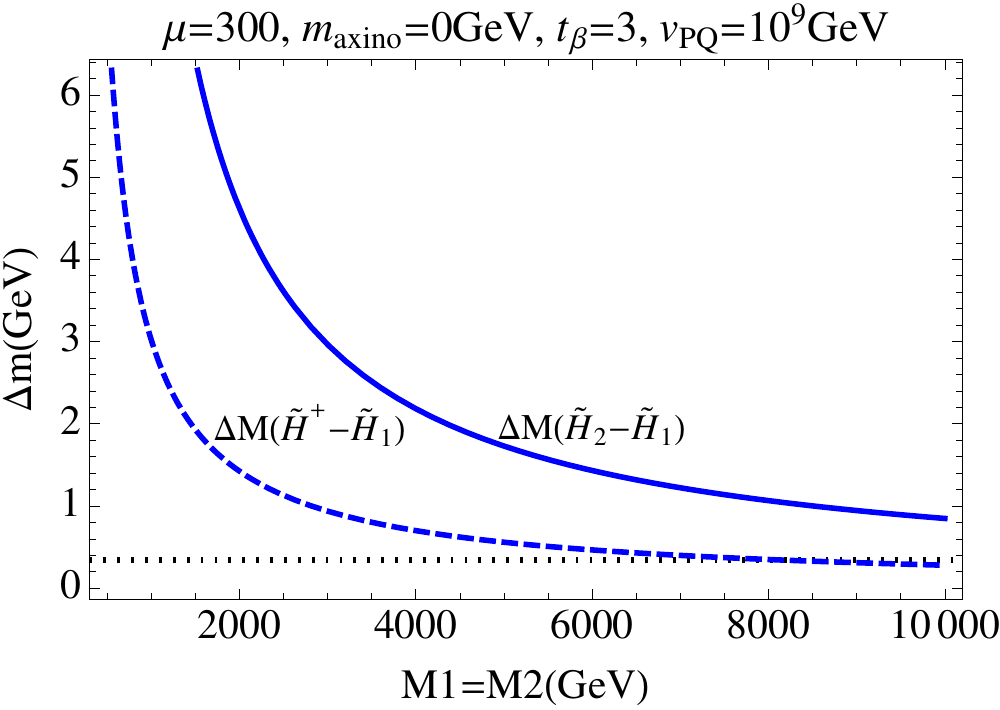}
\caption{\baselineskip=16pt
The decays of heavier Higgsinos (neutral one in the left panel, and charged one in the right panel) to the lightest Higgsino vs. to the axino LSP are compared in the upper panels. Massless axinos and $v_{PQ}=10^9$GeV are assumed here. In the lower panel, we show tree-level mass splittings between Higgsino states. The loop-induced mass splitting of the Higgsinos, $\Delta m \sim 355$MeV~\cite{Thomas:1998wy}, is marked as a horizontal dotted line.
}
\label{fig:H2HpWidth}
\end{figure}

In this section, we consider the situation of light (NLSP) Higgsinos and heavy gauginos with the axino LSP.
In the decoupling limit of gauginos, there occurs an interesting and rich situation for the decays of heavier Higgsinos.
Since the axino LSP is weakly interacting, Higgsinos can dominantly decay either to the lightest Higgsino or to the axino LSP, depending on the gaugino masses and the PQ symmetry breaking scale, $v_{PQ}$. In \Fig{fig:H2HpWidth}, we show relative decay widths of the heavier Higgsinos for massless axinos and $v_{PQ}=10^9$ GeV. Heavier axinos (for a fixed $\mu$) and a higher $v_{PQ}$ scale only make the decays to the axino smaller.
For $M_1=M_2 \lesssim$ a few TeV, both charged and neutral Higgsinos decay dominantly to the lightest Higgsinos even with massless axinos and $v_{PQ}=10^9$GeV. For larger $M_1=M_2$, the mass splitting between Higgsino states are too small to have quick enough decays between them.
In this paper, we simply assume $M_1=M_2=2$TeV for which all heavier Higgsinos decay to the lightest Higgsinos; also as long as $M_1$ and $M_2$ are TeV scales, the mass splitting between Higgsinos are ${\cal O}(1)$GeV(see \Fig{fig:H2HpWidth}) and soft leptons from decays between Higgsinos are too soft to be reliably measurable.

\begin{figure}[t] \centering
\includegraphics[width=0.49\textwidth]{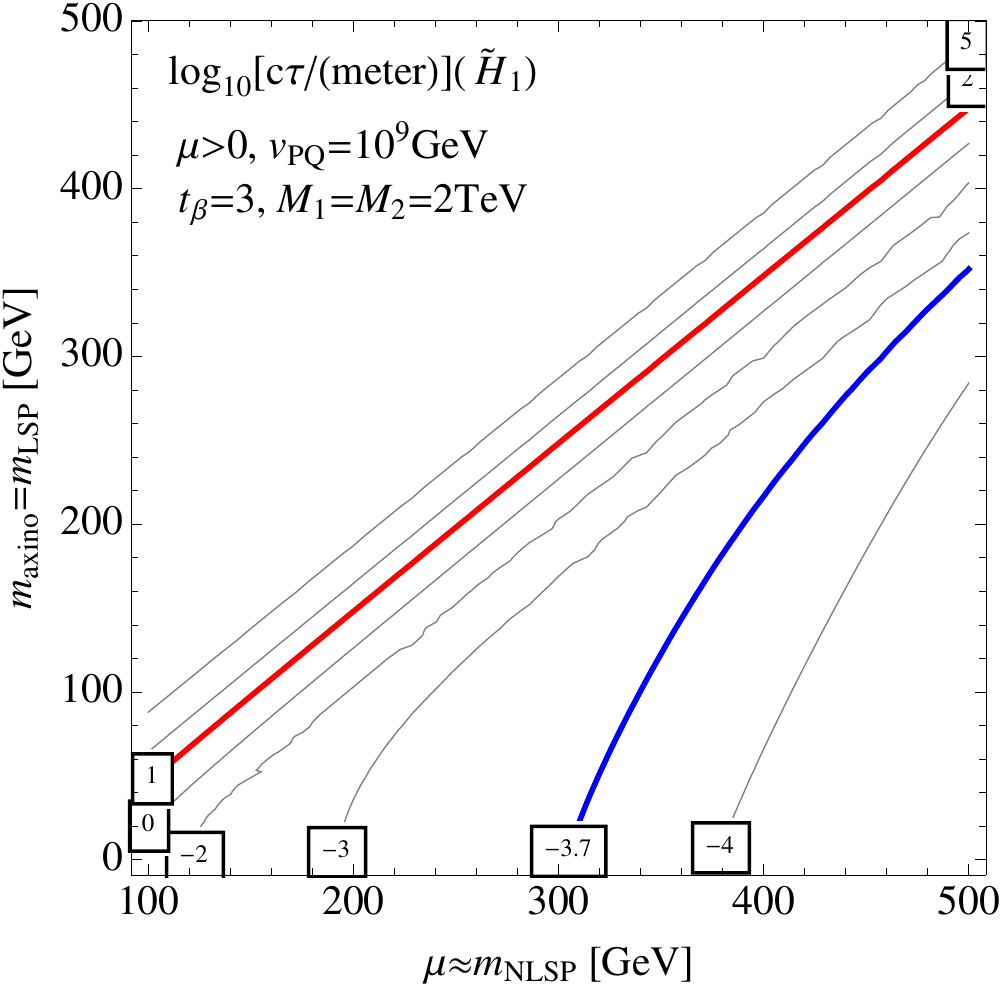}
\caption{\baselineskip=16pt
The proper decay length of the lightest Higgsino NLSP, $\widetilde{H}_1^0$, in the presence of the axino LSP. $v_{PQ}=10^9$GeV here, and the lifetime scales with $v_{PQ}^2$. We mark $c\tau \sim 200\mu$m with a blue line as a convenient reference for the displaced decay, and we mark $c\tau = 10$m with a red line for decaying outside detector. The rapid increase of the lifetime below $m_{\rm NLSP} - m_{\rm LSP} \lesssim 90$GeV is due to the closing of any two-body decay modes.
}
\label{fig:width-higgsino}
\end{figure}

Whether or not the decays of the Higgsino NLSP to the axino LSP can leave observable displaced vertices depends on the values of $\mu$, $v_{PQ}$ and the mass gap between the NLSP and LSP. The proper decay length of the Higgsino NLSP, $\widetilde{H}_1^0$, is shown in \Fig{fig:width-higgsino}.
The distinction between the prompt and displaced decays (also whether decaying inside or outside detector) is not determined solely by the $c\tau$ but also by kinematics of decay products and the probabilistic distributions of decay lengths. But by conveniently referring to the contours of $c\tau = 200\mu$m(blue) and 10m(red) -- standard tight leptons are required to satisfy $d_0 \gtrsim 200\mu $m at LHC~\cite{Aad:2009wy} and the size of ATLAS detector, for example, is $\sim$10m~\cite{Aad:2009wy} --, we find that the decay is most likely be inside detector (and to be displaced) at collider for the favored region of parameter space with a smaller $\mu$ and $v_{PQ} \gtrsim 10^9$ GeV (unless the mass-gap between Higgsino NLSPs and axino LSPs is very small). See Ref.~\cite{Martin:2000eq} for earlier studies of displaced decays of singlinos in a most related context, Refs.~\cite{Meade:2010ji,Graham:2012th} for displaced decays of standard neutralinos and Refs.~\cite{Meade:2009qv,Matchev:1999ft} for lightest Higgsino phenomenology with gravitino LSPs.

\begin{figure}[t] \centering
\includegraphics[width=0.49\textwidth]{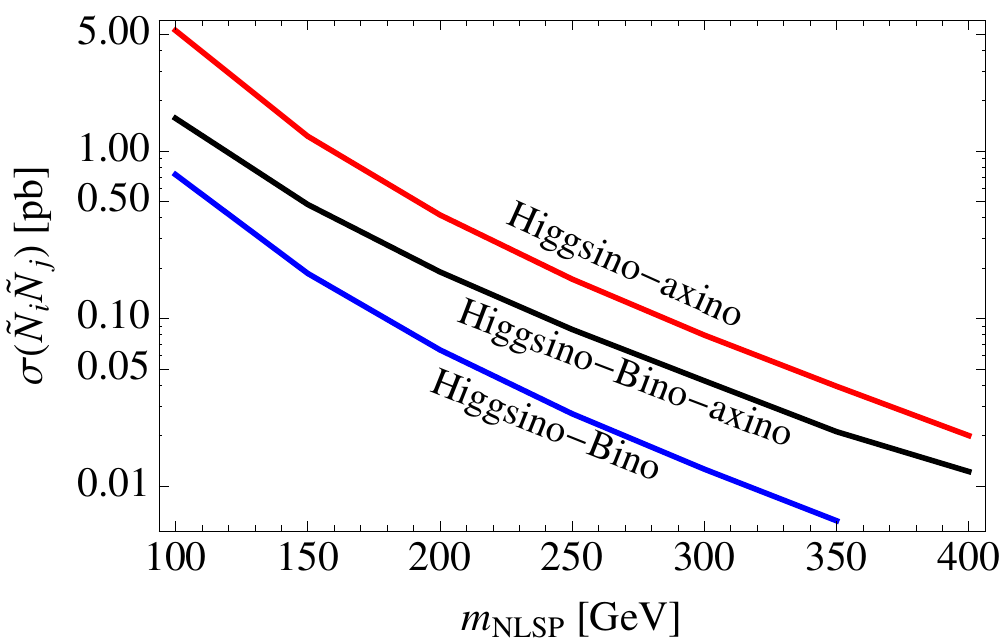}
\caption{\baselineskip=16pt
The production rates of the NLSP neutralino pair effectively relevant to collider physics. For the standard Higgsino-Bino case in \Sec{sec:noaxino}, $\sigma(\widetilde{H}_1^0 \widetilde{H}_2^0)$ is shown as blue. For the Higgsino-axino case in \Sec{sec:axinolsp}, $\sigma(\widetilde{H}_1^0 \widetilde{H}_1^0)$ is shown as red. For the Higgsino-Bino-axino case in \Sec{sec:higgsinonnlsp}, it is the $\sigma(\widetilde{B} \widetilde{B})$ shown as black; $\mu=M_1+50$GeV is assumed. All prompt pair productions of inos effectively leading to the aforementioned production are added; see text for more discussions.}
\label{fig:neutralhiggsinopair}
\end{figure}

\begin{figure}[t] \centering
\includegraphics[width=0.49\textwidth]{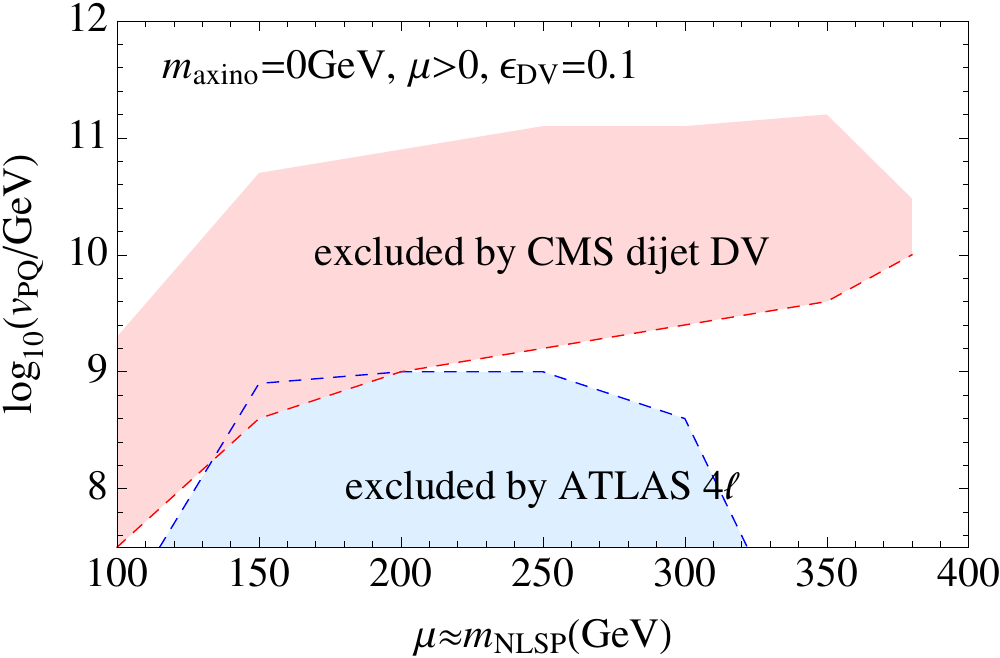}
\includegraphics[width=0.49\textwidth]{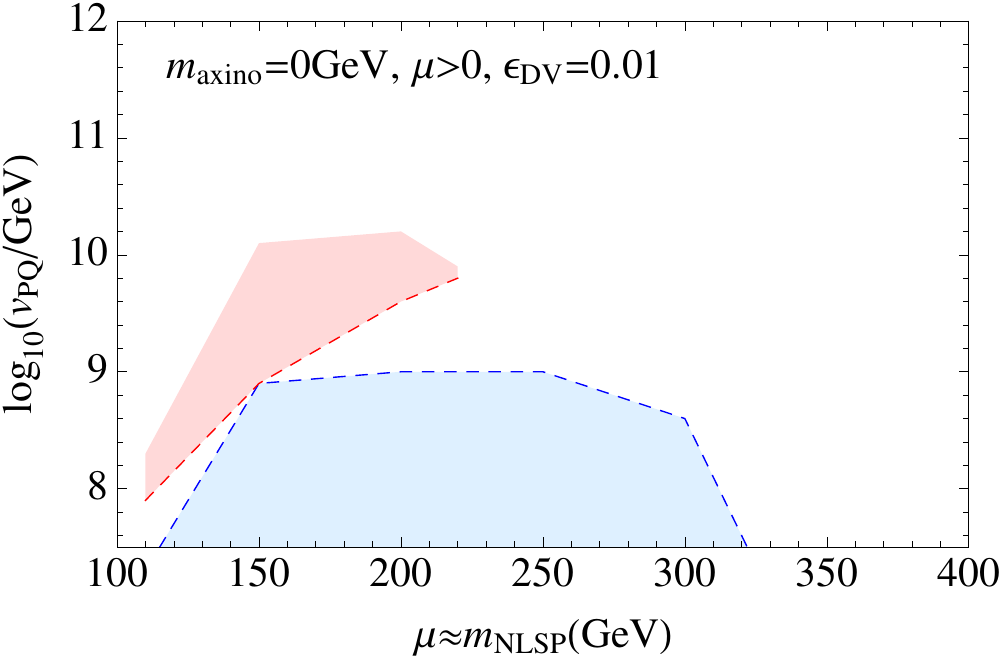}\\
\includegraphics[width=0.49\textwidth]{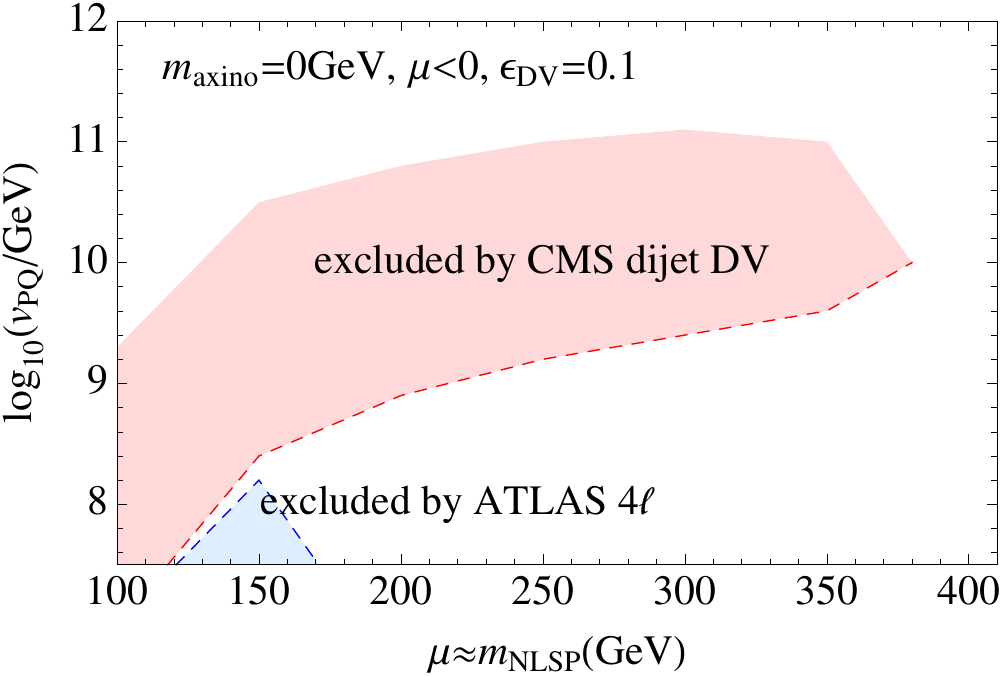}
\includegraphics[width=0.49\textwidth]{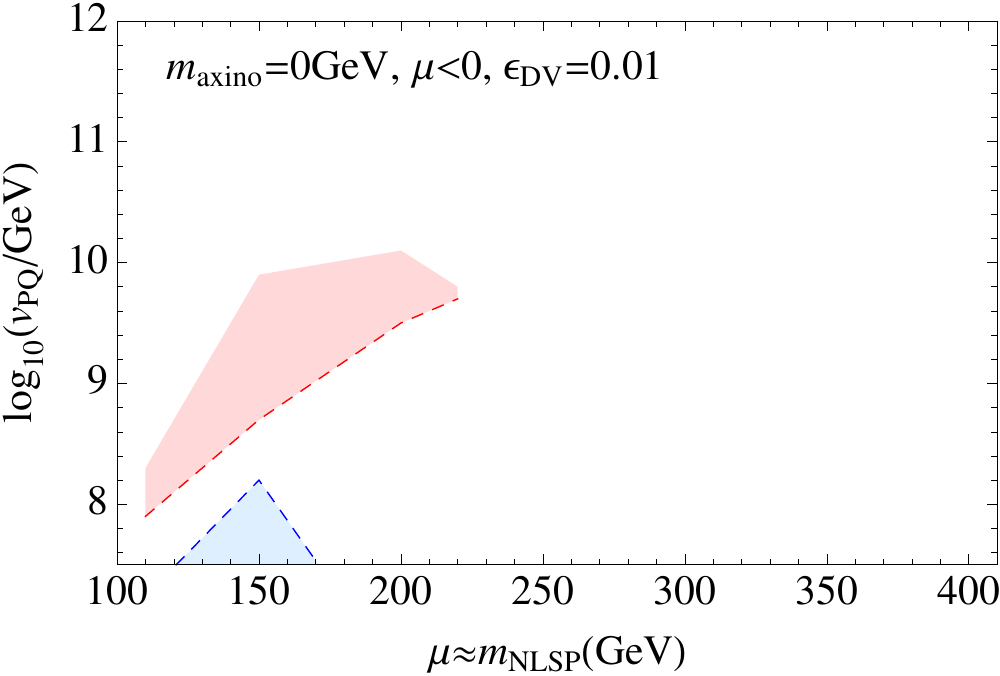}
\caption{\baselineskip=16pt
The excluded parameter space of the case with the Higgsino-NLSP and axino-LSP discussed in \Sec{sec:axinolsp} for $\mu>0$(upper) and $<0$(lower). The $4\ell$+MET search(blue)~\cite{Khachatryan:2014qwa} and the CMS dijet DV search(red)~\cite{CMS:2013oea} are most relevant. We assume two extreme values of the DV reconstruction efficiencies: $\epsilon_{DV}=0.1$(left) and 0.01(right). $m_{ \rm axino} =0$GeV is used, but see \Fig{fig:higgsino-axino2} for results on other values of $m_{\rm axino}$. $M_1=M_2$=2TeV and $t_\beta=3$. More on \Sec{sec:axinolsp}.
}
\label{fig:higgsino-axino1}
\end{figure}

\begin{figure}[t] \centering
\includegraphics[width=0.49\textwidth]{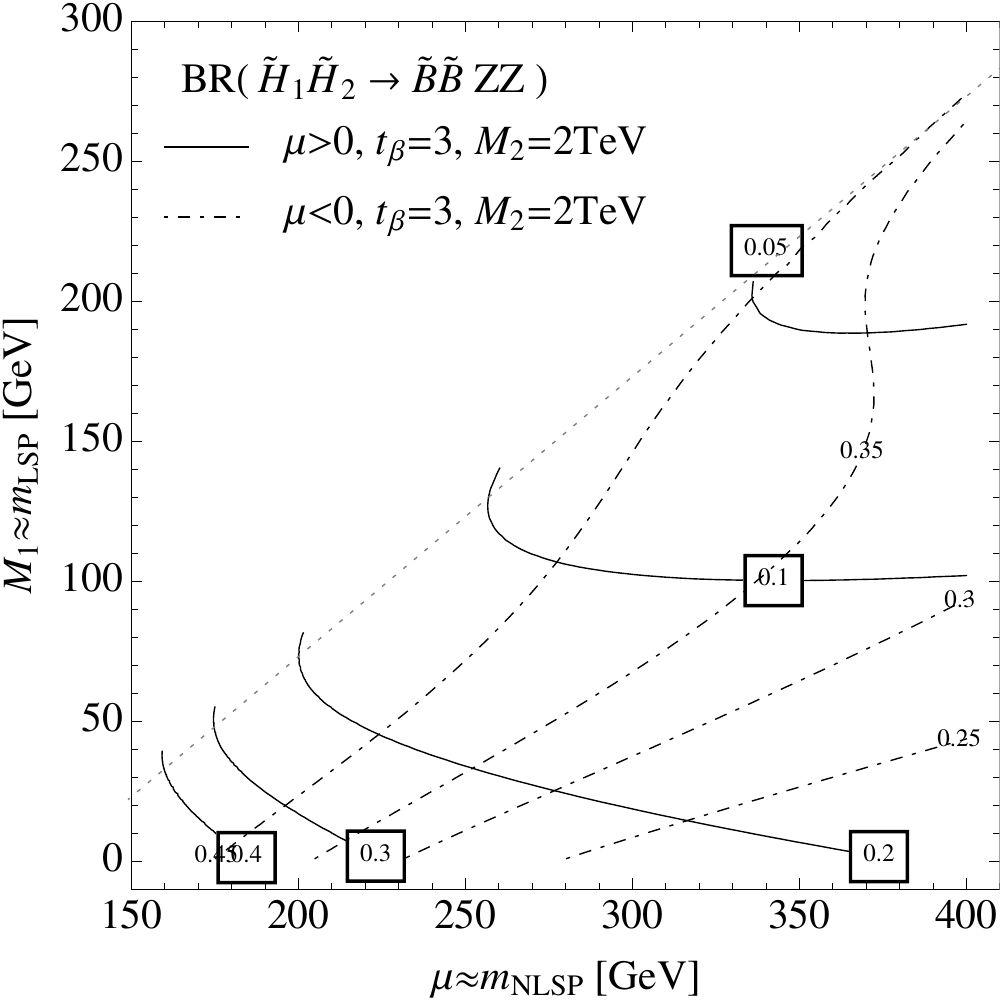}
\includegraphics[width=0.49\textwidth]{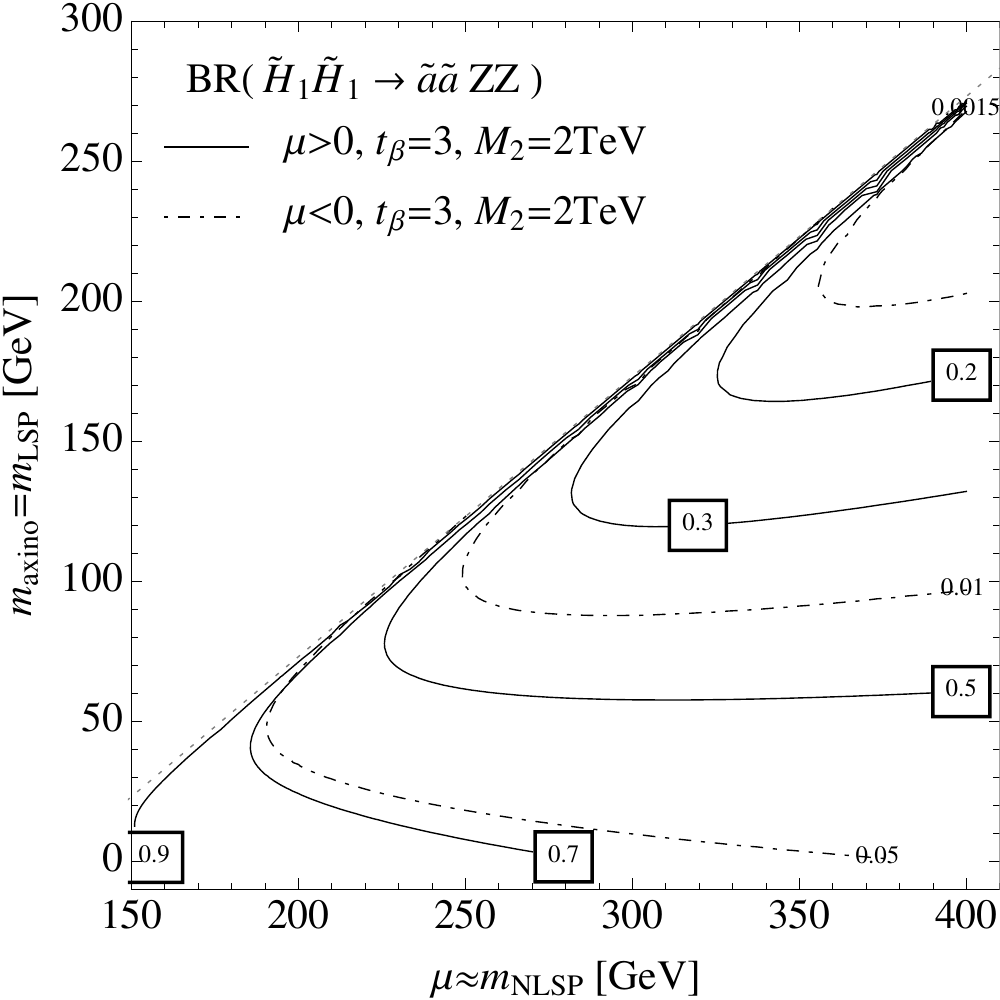}
\caption{\baselineskip=16pt
The BR of the NLSP Higgsino pairs to the $ZZ$ channel relevant to the $4\ell$+MET search. (Left): The Higgsino-Bino case in \Sec{sec:noaxino}. (Right): The Higgsino-axino case in \Sec{sec:axinolsp}. Both $\mu>0$(solid) and $\mu<0$(dashed) are shown. The dotted diagonal lines are at $m_{\rm NLSP} - m_{\rm LSP} \simeq m_h$ above which the decays to the (on- or off-shell) $Z$ boson is almost 100\%. The BR can be larger than a half in the right panel with $\mu>0$. These results are used in obtaining the $4\ell$+MET bounds.
}
\label{fig:H1pairBR}
\end{figure}

Based on the Higgsino decay patterns discussed above, we have a simple scenario where any Higgsino pair productions would essentially be the same as the $\widetilde{H}_1^0 \widetilde{H}_1^0$ pair production and relevant collider signals come only from
$\widetilde{H}_1^0 \to \widetilde a + h/Z$.
It is useful to summarize several differences between the current situation and  the standard Higgsino NLSP and Bino LSP case
discussed in \Sec{sec:noaxino}:
\begin{enumerate}
\item The $\widetilde{H}_1^0 \widetilde{H}_1^0$ production is sizable. Any pair productions of Higgsinos essentially lead to the $\widetilde{H}_1^0 \widetilde{H}_1^0$ and resulting total production rate (adding all) is about 8 times larger than that of the usual $\widetilde{H}_1^0 \widetilde{H}_2^0$ pair production as shown in \Fig{fig:neutralhiggsinopair}. Note that pair productions of neutral Winos or Binos are highly suppressed. The enhanced neutralino pair production can also be resulted in the case with the weakly interacting gravitino LSP~\cite{Meade:2009qv}. 
\item Among standard multi-lepton searches, the $4\ell$+MET search is most relevant through $\widetilde{H}_1^0 \widetilde{H}_1^0 \to \widetilde{a}\, \widetilde{a} \, ZZ \to \widetilde{a} \, \widetilde{a} \, 4\ell$. The $\widetilde{H}_1^0 \widetilde{H}_1^0$ can now contribute to the stringent $3\ell$+MET ($2\ell$+MET as well) searches only by accidentally losing one or more leptons. Thus, such multi-lepton searches are weakened. 
\item Higgsino phenomenology depends only on the decay pattern of $\widetilde{H}_1^0$. Decays of a single neutral Higgsino, $\widetilde{H}_1^0$, depends sensitively on $t_\beta$ and the sign of $\mu$ (as can be seen, e.g. in \Fig{fig:H1pairBR}). On the other hand, in the standard case without axino LSPs, decays of all Higgsino states are indistinguishable at collider and are equally important, and summing all indistinguishable decays make some standard Higgsino phenomenology less sensitive to those parameters~\cite{Jung:2014bda}; see one example in \Fig{fig:BR-higgsinopair}.
\item As discussed, the decay of $\widetilde{H}_1^0$ is likely displaced. The displaced decay further weakens the standard multi-lepton SUSY searches. However, dedicated displaced vertex(DV) searches are now relevant.
\end{enumerate}

\begin{figure}[t] \centering
\includegraphics[width=0.49\textwidth]{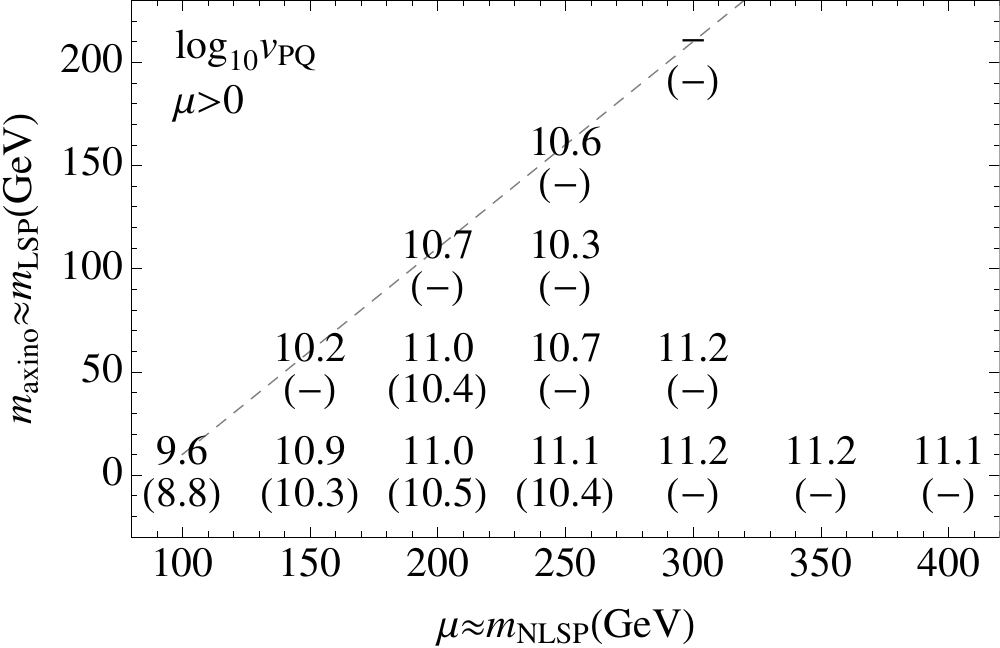}
\includegraphics[width=0.49\textwidth]{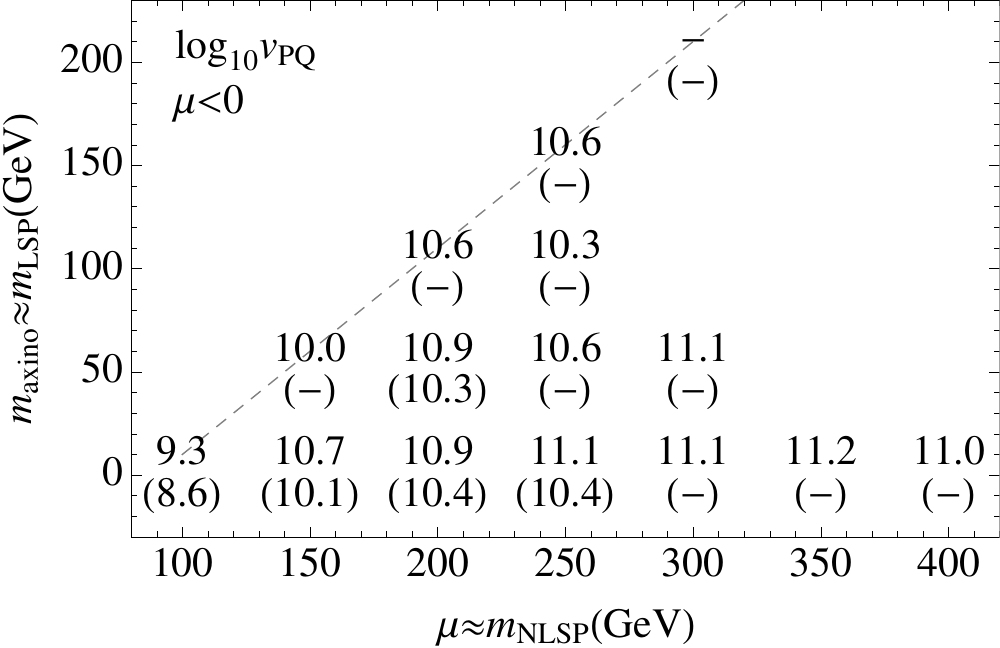}
\caption{\baselineskip=16pt
The highest excluded value of $\log_{10} v_{PQ}$ for the given parameter space of Higgsino NLSP with axino LSP; for example, see \Fig{fig:higgsino-axino1} that the value would be $\sim 11.1$ for the 250-0 case with $\mu <0$. Although there can be a smaller $v_{PQ}$ not excluded, we conveniently choose this highest excluded value to show in these plots. CMS dijet DV and $4\ell$+MET are used. Numbers without(with) parentheses are results with $\epsilon_{DV}=$0.1(0.01). The ``--'' implies no existing bounds. The parameter space without anything written is not simulated by ourselves. The light-gray-dashed diagonal lines imply $m_{\rm NLSP} - m_{\rm LSP} = 90$GeV below which the decays to the off-shell $Z$ boson begins to be phase-space suppressed and $v_{PQ} \gtrsim 10^8$GeV is already high enough to make all Higgsinos decay far outer region or outside the detector -- thus, no collider bounds in general.}
\label{fig:higgsino-axino2}
\end{figure}

In \Fig{fig:higgsino-axino1}, we analyze the exclusion bounds on the $\mu$-$v_{PQ}$ parameter space with $m_{\rm axino}=0$ GeV. Both the $4\ell$+MET search~\cite{Khachatryan:2014qwa} (constraining too much prompt decays) and the CMS dijet DV search~\cite{CMS:2013oea} (constraining a certain range of displaced decay) are relevant. \Fig{fig:higgsino-axino2} shows results in the more general parameter space. For high enough $v_{PQ}$ scales, no bound exists; either the Higgsino decays still dominantly inside detector but its DV is not searched efficiently or the Higgsino dominantly decays outside detector and its phenomenology is essentially the same as that of the Higgsino-LSP case whose current null bounds are discussed in \Sec{sec:noaxino}. The bound from the DV search is sensitive to the DV reconstruction efficiency, $\epsilon_{DV}$, which is an experimental factor capturing how much fraction of DVs are really reconstructed. For the low extreme value of $\epsilon_{DV}=0.01$ (see Ref.~\cite{CMS:2013oea} that $\epsilon_{DV}= 0.01-0.1$ is a reasonable range to consider), the bound almost disappears. The bound from the $4\ell$+MET search is stronger for $\mu>0$ than $\mu<0$ because the relevant BR is larger as depicted in the right panel of \Fig{fig:H1pairBR}. The total decay width of the Higgsino depends slightly on the sign of $\mu$, thus so does \Fig{fig:higgsino-axino2}.

It is useful to understand why the $3\ell$+MET search is now significantly weaker than the $4\ell$+MET search here as opposed to the results of \Sec{sec:noaxino}. The main reason why the $4\ell$+MET search is now sensitive to this model while it is not sensitive to the standard Higgsino-Bino case in \Sec{sec:noaxino} is the enhanced neutral Higgsino pair production in this model as discussed in regard of \Fig{fig:neutralhiggsinopair}. Another minor reason is that the relevant BR (right panel of \Fig{fig:H1pairBR}) can be somewhat larger than a half while it is typically not in the standard case (left panel of \Fig{fig:H1pairBR})
\footnote{The difference is that the same Higgsino, $\widetilde{H}_1^0$, is pair produced here. The decays of $\widetilde{H}_1^0$ and $\widetilde{H}_2^0$ are typically opposite~\cite{Jung:2014bda}.}.
On the other hand, compared to the dominant $\widetilde{H}^\pm \widetilde{H}_1^0$ production in the Higgsino-Bino case leading to the $3\ell$+MET signal, the $\widetilde{H}_1^0 \widetilde{H}_1^0$ here is not much larger, thus a small selection efficiency to the $3\ell$+MET here (needing to accidentally lose one lepton) has a big impact to decrease the exclusion reach of the $3\ell$+MET in this model.

The CMS dilepton DV search~\cite{CMS:2014mca} can also give a relevant bound, but this search looks for a similar range of decay length $\sim 30-60$cm; so we conservatively use the dijet DV results to obtain bounds. Other dedicated DV searches~\cite{Abazov:2009ik} are less relevant and less stringent.

In all, by having the axino LSP, some ranges of $\mu$ and $v_{PQ}$ can be probed at the LHC since the currently allowed range of $v_{PQ}$ falls in the right range to allow NLSP Higgsinos to decay inside detector either promptly or with DVs. On the other hand, a higher $v_{PQ} \gtrsim 10^{10} -10^{11}$GeV with $\mu \sim$ 100-400GeV can avoid all the current LHC searches.
When the mass gap between the Higgsino and the axino is smaller than about $m_Z$, the Higgsino generally decays far outer region or outside the detector and no current collider searches constrain the model.

\section{Higgsino NNLSP, Bino NLSP and axino LSP} \label{sec:higgsinonnlsp}

\begin{figure} \centering
\includegraphics[width=0.49\textwidth]{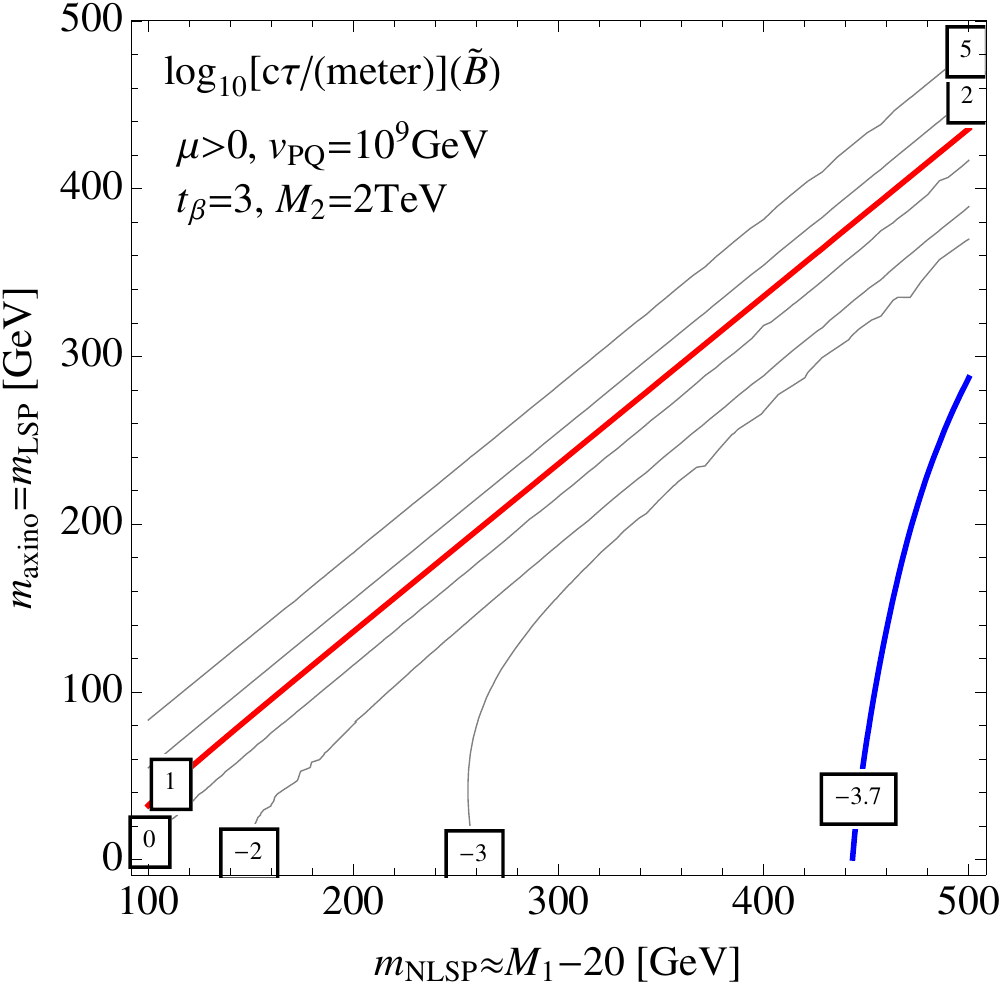}
\caption{\baselineskip=16pt
The proper decay length of the Bino NLSP in the presence of the axino LSP. Figure details are as in \Fig{fig:width-higgsino}. The Higgsino NNLSP is assumed to be nearby with $|\mu| = M_1 + 50 >0$. For a positive $\mu$ assumed here, the mass eigenvalue of the Bino NLSP is related as $m_{\rm NLSP} \simeq M_1 -20$GeV due to the mild mixing between Binos and Higgsinos. More on \Sec{sec:higgsinonnlsp}.
}
\label{fig:width-bino}
\end{figure}

Having another gauginos in the light spectrum is another interesting possibility. In this section, we consider the case with the Higgsino NNLSP, Bino NLSP and axino LSP. As direct Bino pair production is very small, the collider phenomenology relies on all possible pair productions of NNLSP Higgsinos and NLSP Binos. We assume that $|\mu| = M_1 + 50$GeV so that these productions are big enough for collider analysis. Due to this close-by masses and resulting mild mixing between Binos and Higgsinos, the mass eigenvalue of the Bino-like LSP is 20GeV lighter than the $M_1$: $m_{\rm NLSP} \simeq M_1 - 20$GeV. Similarly to the case of the Higgsino NLSP with axino LSP discussed in \Sec{sec:axinolsp}, the produced Higgsinos dominantly and promptly decay to the Bino NLSP. It is more obviously true here because the decays between Higgsinos and Binos are not small-gap-suppressed. Then, again all the Higgsino productions essentially lead to abundant Bino pair productions in the collider physics point of view.

Binos can also decay to the axino LSP with substantial lifetime. In \Fig{fig:width-bino}, we show the proper decay length of Bino NLSPs. In the majority of relevant parameter space, Binos likely decay inside detector either promptly or with DVs. Compared to the Higgsino NLSP's decay in \Fig{fig:width-higgsino}, Binos have a somewhat longer lifetime because Binos couple to axinos via Higgsino mixtures in the DFSZ model. Numerically, it turns out that the Bino typically has a 3--5 times longer lifetime (with the same other parameters) which implies that about 2 times lower $v_{PQ}$ scale is needed for a similar lifetime. If Higgsinos are much heavier, the Bino decays are much slower with about 10--20 times longer lifetime due to a smaller Bino-Higgsino mixing.

\begin{figure}[t] \centering
\includegraphics[width=0.49\textwidth]{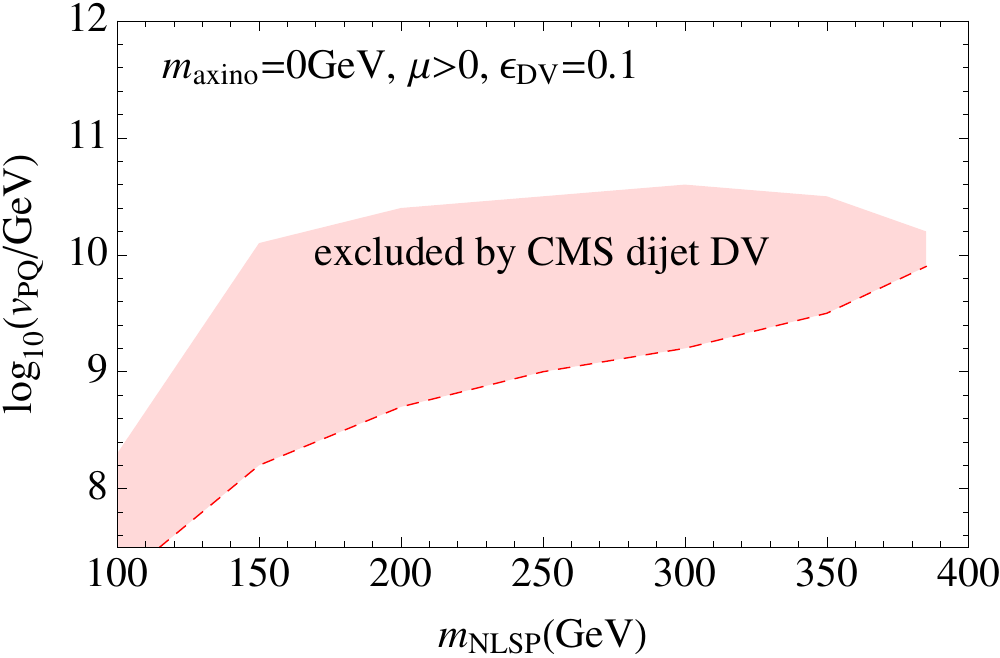}
\includegraphics[width=0.49\textwidth]{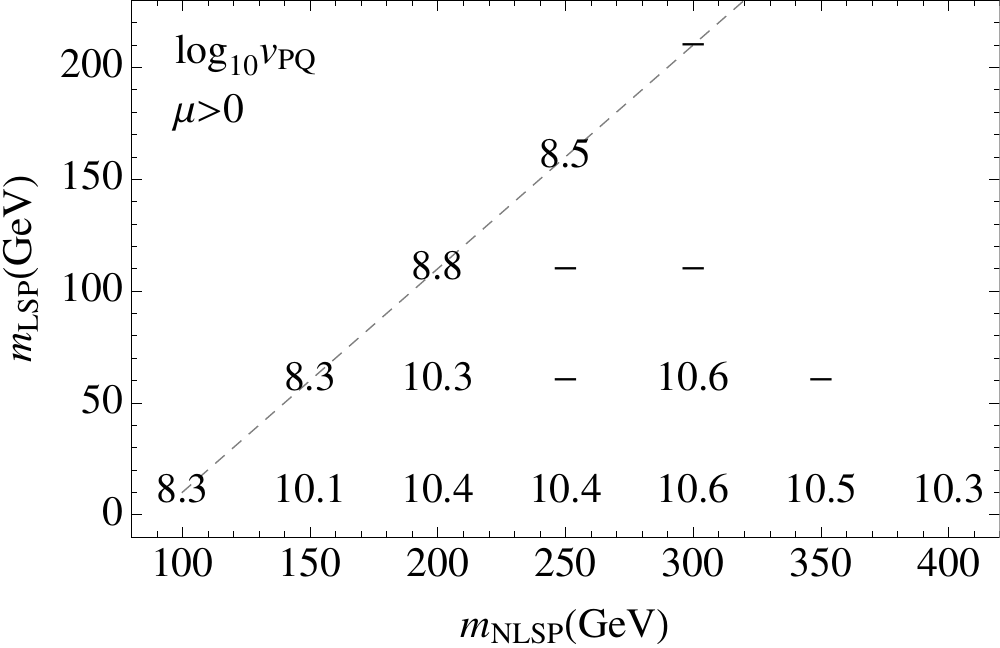}
\caption{\baselineskip=16pt
The highest excluded value of $\log_{10} v_{PQ}$ for the given parameter space of the Higgsino NNLSP, Bino NLSP and axino LSP discussed in \Sec{sec:higgsinonnlsp} with $\mu>0$. Figure details are as in \Fig{fig:higgsino-axino1} and \ref{fig:higgsino-axino2}. No bound is derived from multi-lepton searches in this case. For $\mu<0$, a similar bound is obtained from the DV search. $\epsilon_{DV}=0.1$ here, and no DV bound exists for $\epsilon_{DV}=0.01$.}
\label{fig:BinoDV}
\end{figure}

It is useful to note several differences between this scenario and the Higgsino-axino case in \Sec{sec:axinolsp}. (i) For the given NLSP mass, the effective total production of NLSP pairs is smaller here as shown in \Fig{fig:neutralhiggsinopair} because the model here relies on the (associate) productions of heavier Higgsinos. (ii) Decays of NNLSP Higgsinos to NLSP Binos can produce observable particles as we assume about 50GeV mass-gap. We will explain later how we treat these visible particles in our analysis. (iii) Now the decay pattern of the Bino NLSP is relevant to collider searches instead of that of the Higgsino.

In \Fig{fig:BinoDV}, we analyze the exclusion bounds. Again, both the $4\ell$+MET search (constraining too much prompt decays) and the CMS dijet DV search (constraining a certain range of displaced decay) are relevant. For high enough $v_{PQ}$ scales, no bound exists; either the Bino decays still dominantly inside detector but its DV is not searched efficiently or the Bino dominantly decays outside detector. When the Bino decays outside detector, the visible decay products of NNLSP Higgsinos can be important in collider searches -- the collider physics will then be essentially the same as that of the Higgsino NLSP and Bino LSP considered in \Sec{sec:noaxino} as if axinos were absent. However, the mass-gap between the Higgsino and Bino is only 50GeV in this work, and by referring to \Fig{fig:higgsino3l} showing the current bounds on the Higgsino-Bino model, we find that the visible decay products of Higgsino NNLSP with such small-gap is weakly constrained. We conservatively assume that we can ignore all (soft) leptons from Higgsino decays in our multi-lepton analysis, but we will include all and only leptons from Bino decays to axinos in our analysis (when Binos decay promptly inside detector) -- the more accurate analysis will not give a much stronger bound anyway.

The \Fig{fig:BinoDV}, compared with \Fig{fig:higgsino-axino1} and \ref{fig:higgsino-axino2}, shows that the bound on this model is somewhat weaker than that of the Higgsino-axino case in \Sec{sec:axinolsp}. For $\epsilon_{DV}=0.01$, no bounds from the DV search is derived. For $\mu<0$, no bounds from the multi-lepton search is derived. These weaker bounds are mainly because the effective total production of Bino pairs is smaller for the given Bino mass as discussed above and as shown in \Fig{fig:neutralhiggsinopair}.

The results depend on the choice of $|\mu| = M_1+50$GeV. The heavier Higgsinos, the smaller signal productions and the weaker collider constraints -- it is thus a less interesting scenario. The lighter Higgsino closer to the Bino can induce a larger mixing making Binos decay more promptly (but not faster than pure Higgsinos discussed in previous section) and the excluded parameter space change slightly. If we still assume that lepton from decays between those states are soft enough, not much qualitative change in the collider physics would arise.

But again, in all, by having the axino LSP as well as light gauginos, some ranges of $\mu$ and $v_{PQ}$ can be probed at the LHC since the currently allowed range of $v_{PQ}$ falls in the right range to allow NLSP Binos to decay inside detector either promptly or with DVs. On the other hand, a higher $v_{PQ} \gtrsim 10^{10} -10^{11}$GeV with $\mu \sim$100-400GeV can avoid all the current LHC searches. If the NNLSP Higgsino is much heavier, the model has a looser connection with the naturalness; in any case, no any sizable production modes are available then and the search will rely on heavier particle productions.

\section{LHC14 projection} \label{sec:lhc14}

\begin{figure} \centering
\includegraphics[width=0.49\textwidth]{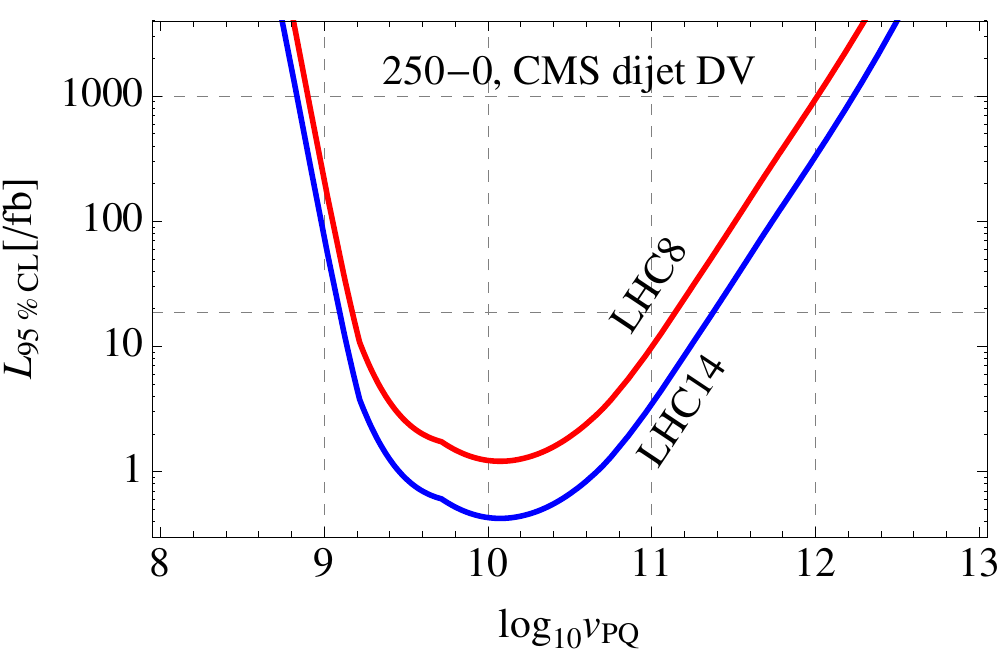}
\includegraphics[width=0.49\textwidth]{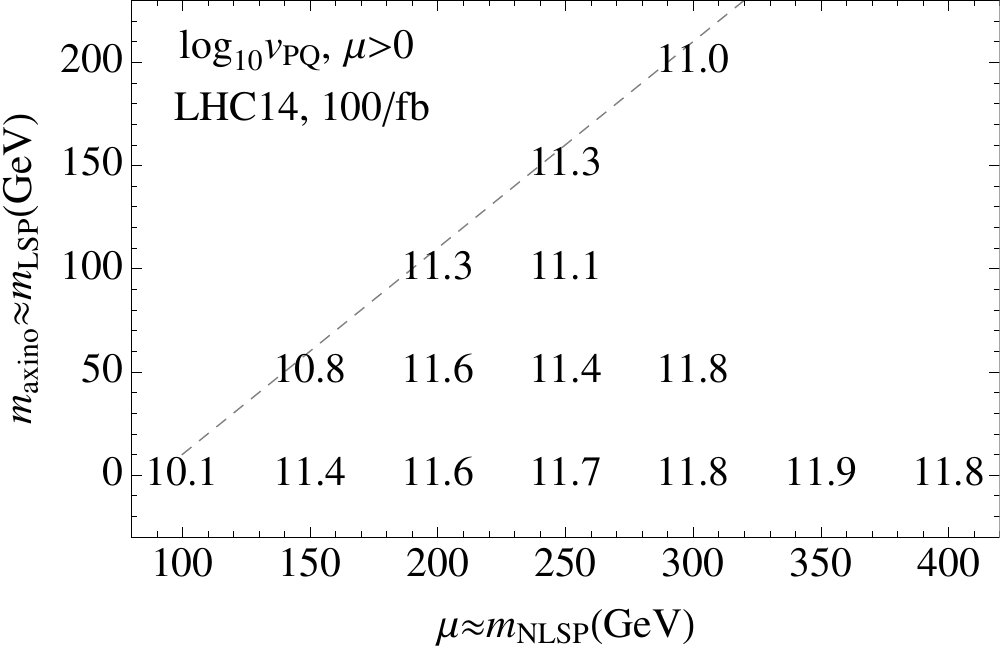}
\caption{\baselineskip=16pt
LHC14 projections of the latest CMS dijet DV search. {\bf Left:} the needed lumonisity for 95\%CL exclusion for the 250-0 parameter space with $\mu>0$ and $t_\beta=3$. The lower horizontal dashed line is at the current CMS data 18.6/fb while the upper horizontal line at 1/ab is just shown for easy reading. {\bf Right:} the highest excluded $\log_{10} v_{PQ}$ value with 100/fb as in the left panel of \Fig{fig:higgsino-axino2} with $\epsilon_{DV}=0.1$. Increasing to 3/ab of data roughly enhances the reach by 0.7 of $\log_{10} v_{PQ}$.
}
\label{fig:higgsino14tev}
\end{figure}

As the LHC 14TeV will start in a year, it is interesting to estimate the prospect of it. We project the current CMS dijet DV search results to study how high $v_{PQ}$ scale can be probed at 14TeV. 

It is a techincally difficult task because future detectors are different and pile-up backgrounds at higher energy collisions are larger. We, however, parameterize the DV reconstruction efficiency which will be most dependent on detector performance by an unknown $\epsilon_{DV}$, and relatively hard cuts on jet $p_T$ used in this analysis ($H_T>300$GeV and $p_T(j)>60$GeV which shall be scaled up at 14TeV) will make the soft pile-up effects less influential.
If we assume that cut/reconstuction efficiencies and the signal-to-background ratio after optimal cuts stay relatively constant between 8TeV and 14TeV analyses, the following simple scaling rule of the statistical significance is obtained\footnote{These are often reasonable assumptions. See Ref.~\cite{Jung:2013zya} where this scaling rule is proven for the search of gluino pairs at future high energy colliders and Ref.~\cite{colliderreach} where a public javascript code can do similar scaling for conventional searches.}
\beq
({\rm significance})_i \= \frac{\sigma_{Si}\, \epsilon_{Si}  \epsilon_{DV} {\cal P}_i}{\sqrt{\sigma_{Bi} \, \epsilon_{Bi}}} \sqrt{ {\cal L}_i} \= \left( \sqrt{\frac{S_i}{B_i} \epsilon_{Si} \epsilon_{DV}} \right) \cdot \sqrt{\sigma_{Si} {\cal P}_i {\cal L}_i },
\ceq
\frac{ ({\rm significance})_i}{({\rm significance)}_j} \= \sqrt{ \frac{ \sigma_{Si} {\cal P}_i {\cal L}_i }{ \sigma_{Sj} {\cal P}_j {\cal L}_j } } \= \sqrt{ \frac{S_i}{S_j} },
\label{eq:scaling} \eeq 
where each factor in the parenthesis, the signal-to-background ratio $S_i/ B_i = (\sigma_{Si} \epsilon_{Si} \epsilon_{DV} {\cal P}_i) / (\sigma_{Bi} \epsilon_{Bi})$, signal cut efficiency $\epsilon_{Si}$ and the assumed $\epsilon_{DV}=0.1$, stay constants as discussed. $\sigma_{Si,Bi}$ are production rates of signal and background, and the probability for displaced decays to be selected by the search, ${\cal P}_i$, depends on the $v_{PQ}$ and mass spectrum. In all, the significance simply scales with the square root of signal event counts. 8TeV CMS dijet DV search bounds can be extrapolated to the 14TeV bounds by finding proper $v_{PQ}$ and mass spectrum giving the same signal event counts as the upper bound of 8TeV results.

We show 14TeV projected results in \Fig{fig:higgsino14tev} obtained in this way. The Higgsino-axino model in \Sec{sec:axinolsp} is used. For the given mass spectrum, LHC14 100(3000)/fb can probe higher $v_{PQ}$ scale by 0.6--0.7(1.3--1.4) of $\log_{10} v_{PQ}$ as shown in the left panel for one choice of parameters. Similar size of improvement is expected for the most of light Higgsino parameter space shown in the right panel. With 3000/fb, $v_{PQ}$ as high as $10^{12}$GeV which is a general upper bound is expected to be probed with light Higgsinos. A more dedicated search will be useful in the near future.


\section{Conclusion} \label{sec:conclusion}

The electroweak-scale axino and Higgsino are perhaps predicted altogether by a naturalness philosophy of particle physics. The implications and the consistency of having both light axinos and Higgsinos are studied in the context of a few benchmark models of supersymmetry. Interestingly, for the typical range of the PQ scale, $10^9 \GeV \lesssim v_{PQ}/N_{DW} \lesssim 10^{12}$GeV,
the electroweak-scale NLSP can still decay to the axino LSP inside detector both promptly and by leaving a DV. The $4\ell$+MET signature from the prompt decay of the NLSP is enhanced among standard SUSY searches as all heavier neutralinos and charginos decay promptly first to NLSP neutralinos so that NLSP neutralino pair productions which are relevant to the collider physics are effectively enhanced. The displaced decay of the NLSP is constrained by dedicated DV searches for a certain range of $v_{PQ}$ typically of $10^9 \lesssim v_{PQ} \lesssim 10^{11}\GeV$ depending on the mass spectrum -- searches for a wider range of decay lengths maybe possible~\cite{Meade:2010ji,Aad:2013txa}. A higher PQ scale of $v_{PQ} \gtrsim 10^{10}-10^{11}\GeV$ with the electroweak-scale $\mu$ or the mass spectrum with small mass-gap between the NLSP and LSP is generally safe from all current collider searches. LHC14, however, is expected to probe the large part of interesting parameter space with light Higgsinos according to our naive estimation, thus a more dedicated search is motivated. We hope that we provided a basic collider physics of the natural supersymmetry with the axino LSP and light Higgsino which can also be complementary to the widely studied axino sector cosmology.

\vspace{8mm}
{\it Acknowledgement.}
S.J. thanks Kiwoon Choi, Hyung Do Kim and Chang Sub Shin for useful conversations. S.J. thanks KIAS Center for Advanced Computation for providing computing resources. 
G.B. acknowledges support from MEC Grant FPA2011-23596, the GV Grant PROMETEOII/2013/017 and
EU FP7 ITN INVISIBLES (Marie Curie Actions, PITN-GA-2011-289442).
E.J.C. was supported in part by the National Science Foundation under Grant No. NSF PHY11-25915.
S.J. is supported in part by National Research Foundation (NRF) of Korea under grant 2013R1A1A2058449.
W.I.P. is supported in part by NRF Research Grant 2012R1A2A1A01006053.

\appendix
\section{Ino decays} \label{app:inodecays}

All the relevant two- and three-body decay widths of inos are calculated and collected in Ref.~\cite{Jung:2014bda} (see also Refs.~\cite{Martin:2000eq,Gunion:1987yh,chun11} for earlier results). In this appendix, we further summarize how we calculate the two-body decays to pions which is relevant when the mass gap is very small $\lesssim {\cal O}(1)$GeV.

The two-body decay $\chi_1^+ \to \chi_1^0 \pi^+$ is calculated as~\cite{Thomas:1998wy,Ibe:2012sx}
\beq
\Gamma(\chi^+ \to \chi^0 \pi^+) \= \Gamma(\pi^+) \cdot \frac{16 \, \delta m^3}{m_\pi m_\mu^2} \left( 1- \frac{m_\pi^2}{\delta m^2} \right)^{1/2} \left( 1- \frac{m_\mu^2}{m_\pi^2} \right)^{-2},
\eeq
where the total decay width of a charged pion is $c\tau = 7.80$m or $\tau = 26.03$ns or $\Gamma(\pi^+) = 2.53\times 10^{-17}$GeV~\cite{Beringer:1900zz}. The mass splitting between the chargino and the neutralino is denoted by $\delta m$. We use $m_\pi = 139.6$MeV, $m_\mu = 105.7$MeV~\cite{Beringer:1900zz}. For $\delta m=164.4$(355)MeV which is the one-loop asymptotic Wino(Higgsino) mass splitting~\cite{Thomas:1998wy,Ibe:2012sx}, the proper decay length is $c\tau = 5.9$(0.34)cm (equivalently, $\tau =$0.20(0.011)ns). The current disappearing track search~\cite{TheATLAScollaboration:2013bia} is sensitive to $\tau \gtrsim 0.1$ns, thus is currently not so sensitive to the nearly degenerate Higgsinos.

\section{Bound estimation} \label{app:simulation}

We list methods and numerical results that we used to obtain various exclusion bounds in this paper. For the $3\ell$+MET result, we use the reported upper limits on the number of events in various \texttt{SR0$\tau$a} bins of Ref.~\cite{Aad:2014nua}. The \texttt{SR0$\tau$a-bin16} is usually strongest for heavy NLSPs. For the $4\ell$+MET result, we interpret the result in the bin of \texttt{2OSSF + 0$\tau_h$} with MET$>$100GeV of Ref.~\cite{Khachatryan:2014qwa} to the upper limit of number of events $N \lesssim 2.0$ at 1.96$\sigma \simeq 95\%$CL. Interestingly, a very similar analysis has been carried out by ATLAS in Ref.~\cite{ATLAS:2013qla}, but their weaker cut on MET$>$75GeV leads to a much weaker bound. Thus, the optimization of the $4\ell$+MET cuts in each parameter space as roughly done for the $3\ell$+MET above will be useful. For the dijet DV result in Ref.~\cite{CMS:2013oea}, we conservatively use the result for $L_{xy}<20$cm (combined with 2 observed events) to obtain the upper limit on the new physics contribution $N\lesssim 3.1$ at 1.96$\sigma \simeq 95\%$CL. For all results, we generate \texttt{MadGraph}~\cite{Alwall:2011uj} events with up to one additional parton and showered them by interfacing with \texttt{pythia}~\cite{Sjostrand:2006za} using the \texttt{MLM}~\cite{Mangano:2006rw} matching. We use \texttt{FastJet}~\cite{Cacciari:2011ma} for particle reconstruction.


\end{document}